\documentclass[runningheads]{llncs}

\usepackage{graphicx}
\usepackage{booktabs}
\usepackage{array}
\usepackage{multirow}
\usepackage{placeins}
\usepackage[tbtags]{amsmath}
\usepackage{amssymb}
\usepackage{xcolor}

\usepackage{setspace}
\usepackage{tikz}
\usetikzlibrary{
  arrows.meta,
  positioning,
  shapes.geometric,
  fit,
  calc,
  shadows
}

\newcolumntype{L}[1]{
  >{\raggedright\arraybackslash}p{#1}
}

\newcolumntype{N}[1]{
  >{\raggedright\arraybackslash
  \hyphenpenalty=10000
  \exhyphenpenalty=10000}p{#1}
}

\usepackage[
    colorlinks=true,
    linkcolor=blue,
    citecolor=green,
    urlcolor=blue
]{hyperref}

\usepackage{xspace}
\newcommand{\BfPara}[1]{\smallskip\noindent\textbf{#1.}\xspace}

\newcommand{\ItPara}[1]{\smallskip\noindent{\em #1.}\xspace}

\usepackage{times}

\definecolor{pal1}{HTML}{E3F2FD}
\definecolor{pal2}{HTML}{BBDEFB}
\definecolor{pal3}{HTML}{90CAF9}
\definecolor{pal4}{HTML}{64B5F6}
\definecolor{pal5}{HTML}{42A5F5}
\definecolor{pal6}{HTML}{2196F3}
\definecolor{pal7}{HTML}{1E88E5}
\definecolor{pal8}{HTML}{1976D2}
\definecolor{pal9}{HTML}{1565C0}
\definecolor{pal10}{HTML}{0D47A1}

\definecolor{pal1deep}{HTML}{2196F3}
\definecolor{pal4deep}{HTML}{1976D2}
\definecolor{pal7deep}{HTML}{1565C0}
\definecolor{pal10deep}{HTML}{0D47A1}

\definecolor{blueaccent}{HTML}{A3C4F3}
\definecolor{stageblue}{HTML}{FBF8CC}
\definecolor{greenaccent}{HTML}{90DBF4}
\definecolor{stagegreen}{HTML}{FDE4CF}
\definecolor{orangeaccent}{HTML}{F1C0E8}
\definecolor{stageorange}{HTML}{FFCFD2}
\definecolor{purpleaccent}{HTML}{CFBAF0}
\definecolor{stagepurple}{HTML}{98F5E1}
\usepackage{pifont}

\definecolor{fsFill}{HTML}{E0F2F1}
\definecolor{fsDraw}{HTML}{00695C}
\definecolor{cotFill}{HTML}{EDE7F6}
\definecolor{cotDraw}{HTML}{512DA8}

\usepackage{tabularx}

\begin{document}

\title{Learning from Failures: A Failure-Driven Prompt Refinement for LLM-Based Vulnerability Analysis}

\author{Mandana Ghadamian \and David Mohaisen}

\authorrunning{M. Ghadamian and D. Mohaisen}

\institute{
University of Central Florida, Orlando, FL 32816, USA\\
\email{\{mandana.ghadamian,mohaisen\}@ucf.edu}
}


\titlerunning{Learning from Failures}

\maketitle

\begin{abstract}

Large Language Models have emerged as promising tools for software vulnerability analysis, but their effectiveness depends heavily on prompt design. Existing research primarily compares prompting strategies using aggregate performance metrics, providing limited insight into why models fail or how prompts can be improved systematically. We propose \textbf{Failure-Driven Prompt Refinement (FDPR)}, a methodology that analyzes recurring model failures to guide evidence-based prompt refinement. Using the Damn Vulnerable Java Application (DVJA), we identify recurring failure modes, including false positives, false negatives, unsupported reasoning, and CWE misclassification, and translate them into targeted prompt refinements. We then evaluate the resulting prompt on the Juliet Test Suite and perform cross-model validation to assess generalizability. The results show that failure-driven refinement improves the reliability of LLM-based vulnerability analysis while yielding reusable prompt design principles. More broadly, this work demonstrates that recurring model failures provide a principled foundation for prompt engineering, enabling the systematic development of more reliable LLM-based vulnerability analysis systems.
\end{abstract}

\section{Introduction}

Software vulnerabilities remain a primary cause of security breaches \cite{pearce2025asleep,jaffal2025large}, making accurate and scalable vulnerability detection a fundamental challenge in secure software development. Traditional static and dynamic analysis techniques improve software assurance but often fail to detect vulnerabilities that require semantic understanding, complex program context, or developer intent \cite{yang2026context}. Recent advances in LLMs have shown strong capabilities in program comprehension, code generation, debugging, and vulnerability analysis, suggesting that they can complement conventional techniques by reasoning about source code using learned programming knowledge \cite{sheng2025llms,camarato2026promptaudit,kharma2026enhancing}. Consequently, LLM-assisted vulnerability analysis has become an active research area.

Despite this progress, deploying LLMs in security-critical workflows remains challenging \cite{kaniewski2026systematic}. Vulnerability analysis requires recognizing security-relevant patterns and producing evidence-based conclusions supported by the analyzed program \cite{sayagh2025think}. In practice, LLMs hallucinate vulnerabilities, generate unsupported reasoning \cite{ullah2024llms}, misclassify Common Weakness Enumeration (CWE) categories \cite{sayagh2025think}, and overgeneralize from superficial code patterns \cite{evertz2025chasing}. These limitations reduce analyst trust, increase verification effort, and hinder the practical adoption of LLM-based vulnerability analysis.

Recent work has explored prompt engineering strategies for software vulnerability detection, including zero-shot prompting, few-shot prompting, role-based instructions, structured output constraints, and domain-specific context \cite{kaniewski2026systematic}. These studies consistently demonstrate that prompt design substantially affects model performance \cite{yang2026context,ceka2024can}. However, prompt refinement remains largely empirical. Existing approaches typically compare prompt variants using aggregate metrics such as accuracy, precision, recall, and F1-score \cite{bae2024enhancing}. While these metrics quantify overall performance, they provide limited insight into why models fail. Consequently, false positives, false negatives, unsupported reasoning, hallucinated evidence, and incorrect CWE assignments are treated as evaluation outcomes rather than engineering signals for systematic prompt refinement.

This limitation is reflected in inconsistent findings. Studies of prompting for software vulnerability detection report conflicting results for few-shot prompting, role prompting, and contextual guidance \cite{kharma2026empirical}. Some report substantial gains from few-shot prompting \cite{dozono2026large}, whereas others find marginal improvements or degraded performance relative to zero-shot prompting \cite{lin2025large}. Similar inconsistencies arise for other prompting strategies \cite{ceka2024can,bae2024enhancing}. These results suggest that prompt effectiveness depends on interactions among the model, dataset, task, and evaluation method, limiting the generalizability of benchmark-based guidelines \cite{kaniewski2026systematic}. We therefore ask three research questions: \textbf{RQ1:} Can recurring LLM failure patterns be systematically identified and organized to reveal prompt weaknesses in software vulnerability analysis? \textbf{RQ2:} Can iterative failure-driven prompt refinement improve the reliability and accuracy of LLM-based vulnerability detection? \textbf{RQ3:} What prompt design principles emerge from recurring failures, and do they generalize beyond the prompt-development dataset?

To answer these questions, we propose \textbf{Failure-Driven Prompt Refinement (FDPR)}, a systematic, evidence-driven methodology for prompt engineering. FDPR identifies recurring failures, infers their causes, organizes them into a taxonomy, prioritizes them by their impact on analyst trust and system reliability, and translates them into targeted prompt revisions. Iterative application of this process produces reusable prompt design principles that improve the reliability of LLM-based vulnerability analysis.

We evaluate FDPR for Java vulnerability detection using a two-stage design. We first refine prompts on the Damn Vulnerable Java Application (DVJA), enabling detailed analysis of model behavior and recurring failure patterns. We then evaluate the refined prompt on the Juliet Test Suite, a substantially larger benchmark spanning diverse CWE categories, to assess generalization beyond the development dataset. Our results show that FDPR substantially reduces unreliable vulnerability reports and shifts the remaining errors from unsupported reasoning to taxonomy-level classification errors. The contributions of this paper are as follows:\vspace{-2mm}

\begin{enumerate}
\item \textbf{Failure-Driven Prompt Refinement.} We introduce FDPR, a systematic methodology for improving prompts through iterative analysis of recurring LLM failures. FDPR combines structured failure analysis with evidence-driven prompt refinement, translating recurring failure patterns into targeted prompt modifications.

\item \textbf{Empirical Validation.} We develop FDPR using the Damn Vulnerable Java Application (DVJA) and evaluate the resulting prompt on the Juliet Test Suite. The results demonstrate that systematic failure analysis improves the reliability of LLM-based vulnerability analysis while generalizing beyond the development benchmark.

\item \textbf{Reusable Design Principles.} We derive reusable prompt design principles, including evidence-based reasoning, context-aware analysis, explicit uncertainty handling, and structured vulnerability classification. Although for software vulnerability, these principles may generalize to other LLM-based software engineering tasks.
\end{enumerate}

\section{Background}
\label{sec:background}

Prompt engineering is the primary approach for adapting LLMs to software vulnerability detection without modifying model parameters \cite{zhang2024prompt}. Rather than fine-tuning, prompts guide models through natural-language instructions and contexts during inference. Prompting techniques have evolved from simple task descriptions to structured reasoning, demonstrations, and multi-stage workflows that improve vulnerability detection. Fig.~\ref{fig:prompt-evolution} summarizes the evolution of the three paradigms we consider.

\definecolor{zs}{HTML}{42A5F5}
\definecolor{fs}{HTML}{1E88E5}
\definecolor{cot}{HTML}{1565C0}
\begin{figure*}[t]
\centering
\resizebox{\textwidth}{!}{
\begin{tikzpicture}[
x=1cm,y=0.4cm,
>=Stealth,
dot/.style={circle,fill,inner sep=1.8pt},
lbl/.style={font=\small,align=left},
title/.style={font=\bfseries\large},
lane/.style={font=\bfseries\normalsize},
]

\draw[thick,-{Stealth[length=3mm]}] (0,0)--(16,0);
\foreach \x/\yr in {0/2020,,15.5/2026}{
\draw (\x,0.08)--(\x,-0.08);
\node[above,font=\normalsize] at (\x,0.1){\yr};
}

\node[lane,text=zs,anchor=east] at (-0.3,-1.4){Zero-Shot (ZS)};
\node[lane,text=fs,anchor=east] at (-0.3,-3.4){Few-Shot (FS)};
\node[lane,text=cot,anchor=east] at (-0.3,-5.4){Chain-of-Thought (CoT)};

\newcommand{\event}[5]{
\node[dot,#4] (#1d) at (#2,#3) {};
\node[lbl,anchor=west] at ($(#1d)+(0.12,0)$) {#5};
}

\event{z1}{0.2}{-1.5}{zs}{\textbf{In-Context Learning}\\Brown et al. (2020)}
\event{z2}{6.0}{-1.5}{zs}{\textbf{ZS Baseline}\\Zhang et al. (2023)}
\event{z3}{10.0}{-1.5}{zs}{\textbf{Role-based Prompts}\\Beri et al. (2023)}
\event{z4}{14.4}{-1.5}{zs}{\textbf{Confidence Estimation}\\Pearce et al. (2025)}
\draw[zs,thick] --(z1d)--(z2d)--(z3d)--(z4d);

\event{f1}{0.2}{-3.5}{fs}{\textbf{FS Learning}\\Brown et al. (2020)}
\event{f2}{3.6}{-3.5}{fs}{\textbf{Core Terminology \& Taxonomy}\\Wang et al. (2020)}
\event{f3}{9.0}{-3.5}{fs}{\textbf{Graph \& In-Context Learn}\\Lu et al. (2024)}
\event{f4}{14.4}{-3.5}{fs}{\textbf{Retrieval \& contrastive}\\Kaniewski et al. (2025)}
\draw[fs,thick] (f1d)--(f2d)--(f3d)--(f4d);

\event{c1}{3.0}{-5.5}{cot}{\textbf{CoT}\\Wei et al. (2022)}
\event{c2}{5.9}{-5.5}{cot}{\textbf{Structured Reasoning}\\Sun et al. (2024)}
\event{c3}{9.9}{-5.5}{cot}{\textbf{Multi-Stage CoT}\\Sheng et al. (2025)}
\event{c4}{13.1}{-5.5}{cot}{\textbf{Multi-Agent}\\Sayagh et al. (2025)}
\event{c5}{16}{-5.5}{cot}{\textbf{Hierarchical CoT}\\Okeke et al. (2026)}
\draw[cot,thick] (c1d)--(c2d)--(c3d)--(c4d)--(c5d);

\end{tikzpicture}
}
\caption{Evolution of zero-shot, few-shot, and chain-of-thought prompting
techniques relevant to LLM-based software vulnerability detection, from
their foundational general-purpose formulations to increasingly
specialized adaptations for this task.}
\label{fig:prompt-evolution}\vspace{-5mm}
\end{figure*}
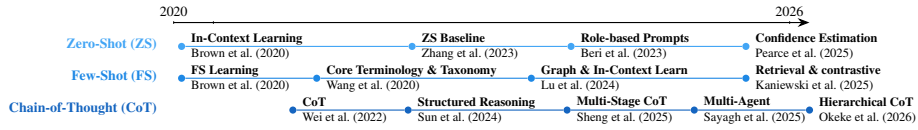

\BfPara{Zero-Shot Prompting} 
Brown et al. showed that LLMs can perform downstream tasks through in-context learning without parameter updates \cite{brown2020language}. Zero-shot prompting applies this capability to vulnerability detection by providing only task instructions and the target program. It has been widely adopted as a baseline across benchmark datasets and commercial LLMs \cite{zhang2024prompt}. More recent work augments zero-shot prompts with role-based instructions, confidence estimation, and security-oriented templates to improve reasoning consistency without demonstrations \cite{pearce2025asleep}.

\BfPara{Few-Shot Prompting} 
Few-shot prompting extends zero-shot prompting by including a small number of demonstrations, allowing models to infer task objectives without additional training \cite{brown2020language}. It is widely recognized as an effective adaptation strategy for foundation models \cite{wang2020generalizing}. In vulnerability detection, demonstrations typically include representative vulnerable and benign code, CWE descriptions, and structured reasoning examples. Recent work has explored graph-based demonstrations, similarity-based retrieval, contrastive vulnerable-patched pairs, and automatically generated examples to improve detection accuracy \cite{kaniewski2026systematic}.

\BfPara{Chain-of-Thought Prompting}
Chain-of-Thought (CoT) prompting improves performance by eliciting intermediate reasoning steps \cite{wei2022chain}. In vulnerability detection, CoT encourages models to analyze program behavior before identifying security weaknesses. Recent work has introduced structured reasoning stages, hierarchical and multi-stage prompting, self-consistency, Tree-of-Thought prompting, and multi-agent collaboration to improve vulnerability localization and CWE classification.

These advances demonstrate the value of structured prompting for LLM-based vulnerability analysis. FDPR builds on this foundation by systematically refining prompts based on recurring failure patterns rather than ad hoc prompt engineering.

\section{Related Work}
\label{sec:related}

\BfPara{Traditional Software Vulnerability Detection} Deep learning shifted vulnerability detection from  engineered features to learned  representations. Representative approaches include token-based models, semantic code gadgets \cite{li2018vuldeepecker}, graph neural networks \cite{zhou2019devign}, program-slicing representations, and Transformer-based models capable of vulnerability localization \cite{fu2022linevul}. While these methods improved accuracy, they rely on fixed learned representations and require retraining to adapt to new vulnerability patterns.

\BfPara{Datasets and Benchmarking} 
Progress in vulnerability detection has been driven by increasingly realistic benchmarks. Recent datasets improve label quality through differential analysis \cite{zheng2021d2a}, support multilingual evaluation, increase project and CWE diversity \cite{chen2023diversevul}, reduce duplication and labeling errors \cite{ding2024vulnerability}, and provide controlled micro-benchmarks \cite{dubniczky2025castle}. These studies show that benchmark quality strongly influences reported performance and that many approaches fail to generalize across datasets.

\BfPara{LLM-Based Vulnerability Detection} 
LLMs have recently been applied to vulnerability classification, localization, and security reasoning \cite{liu2024exploring,khare2025understanding}. Existing studies consistently report that LLMs recognize common vulnerability patterns but remain prone to false positives, unreliable reasoning, weak localization, and inconsistent vulnerability classification \cite{steenhoek2024comprehensive,ullah2024llms}. Prompt engineering, retrieval augmentation, and program-analysis-derived context partially mitigate these limitations \cite{zhang2024prompt,wei2022chain,ceka2024can,sun2024llm4vuln,du2026enhancing,yang2026context}, although their effectiveness varies across models, datasets, and tasks.

\BfPara{Prompt Refinement} 
Most prompt engineering studies compare predefined prompting strategies or optimize prompts using benchmark performance \cite{chen2026promptware}. Although these approaches improve accuracy, they treat model failures as evaluation outcomes rather than diagnostic evidence for refinement. Consequently, they provide limited guidance for systematically improving prompt reliability.

\BfPara{Positioning of FDPR} 
FDPR complements existing prompt engineering by focusing on the refinement process rather than individual prompting strategies. It systematically identifies recurring failure modes, attributes them to specific reasoning deficiencies, and translates them into targeted prompt revisions. Rather than introducing a new prompting technique, FDPR provides an evidence-driven methodology for developing reliable prompts for LLM-based software vulnerability analysis.

\section{Methodology}
\label{sec:Methodology}

\subsection{Overview}

\begin{figure*}[t]
\centering
\resizebox{\textwidth}{!}{
\begin{tikzpicture}[
  font=\sffamily\normalsize,
  node distance=5mm and 5mm,
  every node/.style={align=center},
  stagebox/.style={rectangle, rounded corners=3pt, line width=1pt,
                    text width=2.6cm, minimum height=2.0cm, inner sep=2pt,
                    font=\sffamily\normalsize, text=black, drop shadow},
  arrow/.style={-{Latex[length=2.4mm, width=1.8mm]}, line width=1.3pt},
  decbox/.style={diamond, line width=1.2pt, fill=pal8, draw=pal8!40!black,
                 aspect=1.7, text width=1.9cm, inner sep=1pt,
                 font=\sffamily\bfseries\normalsize, text=white,
                 drop shadow}
]
  \node[stagebox, fill=pal1, draw=pal1!45!black] (init)
    {\textbf{Initial Prompt}};
  \node[stagebox, fill=pal2, draw=pal2!45!black, right=of init] (obs)
    {\textbf{Failure Observation \& Root Cause Analysis}};
  \node[stagebox, fill=pal3, draw=pal3!45!black, right=of obs] (tax)
    {\textbf{Taxonomy Construction \& Prioritization}};
  \node[stagebox, fill=pal4, draw=pal4!45!black, right=of tax] (design)
    {\textbf{Failure-Guided Prompt Refinement}};
  \node[stagebox, fill=pal5, draw=pal5!45!black, right=of design] (reval)
    {\textbf{Prompt Re-evaluation}};
  \node[decbox, right=8mm of reval] (dec) {Dominant failures reduced?};
  \node[stagebox, fill=pal10, draw=pal10!45!black, text=white, right=8mm of dec] (end)
    {\textbf{Extract Design Principles}};
  \draw[arrow, pal1!55!black] (init) -- (obs);
  \draw[arrow, pal2!55!black] (obs) -- (tax);
  \draw[arrow, pal3!55!black] (tax) -- (design);
  \draw[arrow, pal4!55!black] (design) -- (reval);
  \draw[arrow, pal5!55!black] (reval) -- (dec);
  \draw[arrow, pal9!45!black] (dec.east) -- node[above, xshift=-1.5mm, font=\sffamily\bfseries\Large, text=pal9!35!black]{Yes} (end.west);
  \draw[arrow, pal2!55!black] (dec.north) -- ++(0,0.6) -| node[above, pos=0.25, font=\sffamily\bfseries\Large, text=pal2!55!black]{No} (obs.north);
\end{tikzpicture}
}\vspace{-2mm}
\caption{Overview of the FDPR cycle. Each iteration transforms observed failures into prioritized prompt refinements until dominant failure categories are substantially reduced.}
\label{fig:fdpr-workflow}\vspace{-4mm}
\end{figure*}
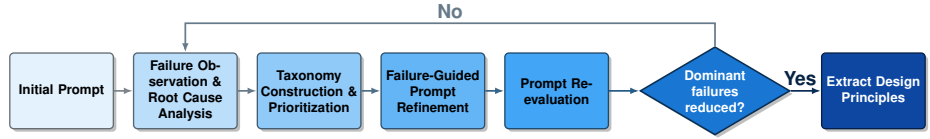

Prompt engineering for vulnerability detection is often guided by aggregate metrics; precision, recall, and F1-score, which quantify performance but provide little insight into why models fail or how prompts should evolve. FDPR instead treats failures as the refinement signal: each failure is analyzed, categorized into recurring failure types, prioritized by its impact, and translated into a targeted prompt refinement. As shown in Fig.~\ref{fig:fdpr-workflow}, FDPR iteratively performs \ding{172} failure observation and root-cause analysis, \ding{173} taxonomy construction and prioritization, \ding{174} failure-guided prompt refinement, \ding{175} prompt re-evaluation, and \ding{176} design principle extraction until further refinements no longer improve performance or the remaining failures reflect inherent model limitations. The resulting prompt is an artifact of the methodology, not its primary contribution.

\subsection{Failure Analysis}
\label{sec:failure-analysis}

\BfPara{Failure Observation}
FDPR begins by systematically analyzing model outputs to identify failures. A failure is any output that violates the requirements of reliable vulnerability detection, including incorrect vulnerability identification, unsupported reasoning, incorrect CWE classification, false positives, false negatives, and malformed responses. Each failure is recorded together with the model output, ground truth, and supporting source-code context to preserve the evidence required for subsequent analysis.

\BfPara{Root Cause Analysis}
Each failure is analyzed to identify the underlying mechanism rather than treating all incorrect outputs as equivalent. Specifically, we determine (1) what the model produced, (2) why it is incorrect, (3) which reasoning mechanism caused the failure, and (4) what prompt modification could prevent it. This distinguishes failures attributable to prompt design from those arising from model limitations, ambiguous program semantics, or the evaluation pipeline.

\BfPara{Failure Taxonomy}
Recurring root causes are grouped into a taxonomy of failure categories based on the reasoning deficiencies that produced them rather than their observable outcomes. This abstraction transforms isolated failures into recurring patterns that can be systematically addressed through prompt refinement.

\BfPara{Failure Prioritization}
Failure categories are prioritized according to their prevalence, security impact, verification cost, recurrence, and expected responsiveness to prompt refinement. Each refinement cycle targets the highest-priority category, ensuring that prompt modifications address the most consequential and actionable failures first.

\subsection{Failure-Driven Prompt Refinement}

Following failure prioritization, FDPR translates the highest-priority failure category into targeted prompt modifications. Rather than relying on intuition or trial and error, each refinement is explicitly linked to a recurring failure mechanism identified during the preceding analysis. \noindent\ding{182} \textbf{Refinement Objective.} The prioritized failure category is analyzed to identify the prompt deficiency responsible for the observed behavior and define a refinement objective that addresses the underlying cause while minimizing unintended effects on unrelated model behavior. 
\noindent\ding{183} \textbf{Prompt Modification.} The refinement objective is translated into one or more prompt modifications, including clarifying task instructions, strengthening reasoning constraints, introducing structured output requirements, or incorporating representative few-shot examples. Every modification is directly traceable to the failure category it is intended to mitigate. 
\noindent\ding{184} \textbf{Iterative Refinement.} Prompt modifications are introduced incrementally and evaluated after each iteration. This isolates the effect of individual changes, enables systematic analysis of their impact, and reduces the risk of introducing new failure modes. The refinement process continues until further modifications yield no meaningful improvement or the remaining failures are attributable to factors beyond prompt design.

\subsection{Iterative Refinement Process}

FDPR follows an iterative refinement process in which each prompt revision is evaluated to determine whether the targeted failure category has been reduced and to identify newly emerging failure modes. After each iteration, the updated prompt is executed on the development dataset, and the resulting outputs undergo the failure analysis described above. This feedback loop ensures that subsequent refinements are driven by empirical observations rather than assumptions about model behavior. 
\noindent\ding{182} \textbf{Incremental Refinement:} Each iteration targets a single prioritized failure category before introducing additional changes. This isolates the effect of individual prompt modifications, enables systematic evaluation of their contribution, and reduces the risk of introducing new failure modes. 
\noindent\ding{183} \textbf{Stopping Criterion:} Refinement continues until further prompt modifications yield no meaningful improvement or the remaining failures are attributable to factors beyond prompt design, such as model limitations or ambiguous program semantics. The resulting prompt is then frozen and evaluated on an independent benchmark.

\subsection{Design Principle Extraction}

FDPR aims to improve an individual prompt and derive reusable principles. Across successive refinement iterations, prompt modifications that consistently mitigate recurring failure categories are abstracted into general design principles rather than retained as prompt-specific adjustments. 
\noindent\ding{182} \textbf{Generalization:} A refinement strategy is promoted to a design principle only if it consistently reduces the targeted failure category across multiple failure instances and refinement iterations. Modifications that improve only isolated cases remain prompt-specific refinements. 
\noindent\ding{183} \textbf{Reusable Knowledge:} The resulting design principles capture the rationale behind successful refinements and provide reusable guidance for prompt development in software vulnerability analysis. By abstracting transferable refinement strategies rather than individual prompt modifications, FDPR produces knowledge that extends beyond a single prompt or benchmark.

\section{Experimental Setup}
\label{sec:setup}

We evaluate FDPR using a two-phase design. Prompt refinement is performed exclusively on the Damn Vulnerable Java Application (DVJA), while the final prompt is frozen and evaluated on the Juliet Test Suite without further modification. This separation prevents benchmark leakage and evaluates whether the refinement methodology generalizes beyond the development environment.

\BfPara{Datasets} We use two datasets. \ding{172} {\em Damn Vulnerable Java Application (DVJA).}  
DVJA serves as the development dataset for iterative prompt refinement. It contains 13 intentionally vulnerable Java programs covering the OWASP Top 10 (2013) categories. Its documented vulnerabilities and CWE mappings enable manual inspection of model outputs, construction of the failure taxonomy, and iterative prompt refinement. \ding{173} {\em Juliet Test Suite.} 
The Java version of the Juliet Test Suite (v1.3) serves as the external validation benchmark. It contains approximately 40,845 Java programs spanning 112 CWE categories. To prevent benchmark leakage, vulnerability annotations, CWE-revealing comments, and identifying file and directory names are removed before evaluation. Ground-truth labels are used exclusively for performance evaluation.

\BfPara{Models} 
Prompt refinement used four locally hosted LLMs: Qwen3-Coder-30B, Qwen3.5-35B, DeepSeek-Coder-6.7B, and Llama-3.2-3B. External validation was performed using Qwen3-Coder-30B and Qwen3.5-35B, the two strongest-performing models during development. All models were served through Ollama with temperature set to 0. Context length, output limits, and quantization used each model's default configuration.

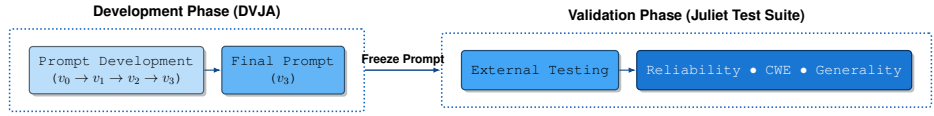
\begin{figure*}[t]
\centering
\resizebox{\textwidth}{!}{%
\begin{tikzpicture}[
  font=\ttfamily\large,
  box/.style={rectangle, rounded corners=3pt, line width=1pt, align=center,
              minimum height=1.15cm, minimum width=3cm, inner sep=8pt,
              font=\ttfamily\large, drop shadow},
  arr/.style={-{Latex[length=2.2mm,width=1.6mm]}, line width=0.9pt, draw=pal9!70!black},
  bigbox/.style={rectangle, draw=pal9, dotted, line width=1.3pt, rounded corners=2pt,
                 inner xsep=16pt, inner ysep=8pt},
  biglabel/.style={font=\sffamily\bfseries\large, text=black}
]

  \node[box, fill=pal2, draw=pal2!45!black, text=black, minimum width=4.6cm, yshift=-0.35cm] (promptdev)
    {Prompt Development\\($v_0 \rightarrow v_1 \rightarrow v_2 \rightarrow v_3$)};
  \node[box, fill=pal4, draw=pal4!45!black, text=black, minimum width=3.0cm, right=0.5cm of promptdev] (finalprompt)
    {Final Prompt\\($v_3$)};

  \draw[arr] (promptdev.east) -- (finalprompt.west);

  \node[bigbox, fit=(promptdev)(finalprompt), inner ysep=14pt] (devpanel) {};
  \node[biglabel, anchor=south] at ([yshift=4pt]devpanel.north) {Development Phase (DVJA)};

  \node[box, fill=pal5, draw=pal5!45!black, text=black, minimum width=3.6cm, right=3.4cm of finalprompt] (extest) {External Testing};
  \node[box, fill=pal8, draw=pal8!45!black, text=white, minimum width=5.8cm, right=0.5cm of extest]
    (reliab) {Reliability $\bullet$ CWE $\bullet$ Generality};

  \draw[arr] (extest.east) -- (reliab.west);

  \node[bigbox, fit=(extest)(reliab), inner ysep=14pt] (valpanel) {};
  \node[biglabel, anchor=south] at ([yshift=4pt]valpanel.north) {Validation Phase (Juliet Test Suite)};

  \draw[arr, line width=1.2pt] (devpanel.east) -- (valpanel.west)
    node[midway, above, font=\sffamily\bfseries\normalsize, text=black] {Freeze Prompt};

\end{tikzpicture}%
}\vspace{-3mm}
\caption{Overview of the prompt development pipeline. The development phase
(DVJA) iteratively refines the prompt through FDPR cycles ($v_0 \rightarrow
v_1 \rightarrow v_2 \rightarrow v_3$) to produce a final prompt, which is
frozen and carried into the validation phase (Juliet Test Suite) for
external testing and evaluation of reliability, CWE coverage, and
generality.}
\label{fig:dev-val}\vspace{-3mm}
\end{figure*}

\BfPara{Experimental Pipeline} 
Fig.~\ref{fig:dev-val} summarizes the evaluation workflow. Prompt refinement iteratively produced four prompt versions (v0--v3) using DVJA. After refinement, the final prompt (v3) was frozen and evaluated on the Juliet Test Suite without modification.

\ItPara{System Prompt} 
The final prompt assigns the model the role of a security auditor and requires evidence-backed source-to-sink reasoning before reporting a vulnerability. It incorporates contrastive vulnerable and non-vulnerable examples, suppresses unsupported findings, and enforces a structured JSON output containing the vulnerability location, CWE, confidence score, and evidence chain (\texttt{input\_source}, \texttt{propagation}, \texttt{sink}, and \texttt{sanitized}). Earlier prompt versions introduced these constraints incrementally throughout the refinement process (Figure~\ref{fig:prompt-versions}).

\ItPara{Input Formatting} 
Each source file is analyzed independently using a single-turn request with file path and source code. No conversation history is retained across files. Juliet programs are preprocessed to remove vulnerability annotations and CWE identifiers, whereas DVJA source files are analyzed without modification.

\ItPara{Prompt Evolution} 
FDPR produced four prompt versions (v0--v3), illustrated in Figure~\ref{fig:prompt-versions}. Each revision addressed the dominant failure category identified during the previous iteration, progressively reducing unsupported findings and improving reasoning consistency. The final prompt (v3) was used throughout external validation.

\begin{figure*}[t]
\centering
\resizebox{\textwidth}{!}{
\begin{tikzpicture}[
    font=\sffamily,
    connector/.style={dashed, line width=0.9pt},
    toolbox/.style={
        fill=white,
        rounded corners=0.18cm,
        minimum width=10.0cm,
        minimum height=0.8cm,
        draw=none
    }
]

\node[font=\sffamily\bfseries\large, black] at (0.0,0.4) {PROMPT VERSION};
\node[font=\sffamily\bfseries\large, black] at (7.6,0.4) {OUTCOME};
\node[font=\sffamily\bfseries\large, black] at (18.0,0.4) {FAILURE-GUIDED DESIGN DECISIONS};

\draw[black!40, line width=0.5pt] (-2.0,0) -- (25.2,0);

\filldraw[fill=pal1, draw=none]
(-2,-0.2) -- (3.5,-0.2) -- (4.5,-1.4) -- (3.5,-2.6) -- (-2,-2.6) -- cycle;
\filldraw[fill=pal1!45!black] (-1.6,-0.5) circle (0.36);
\node[text=white,font=\sffamily\bfseries\large] at (-1.6,-0.5) {v0};
\node[anchor=north west, font=\sffamily\bfseries\large, pal1!40!black] at (-1.2,-0.18){MINIMAL PROMPT};
\node[anchor=north west,align=left,font=\sffamily\large,text width=5.0cm,text=black] at (-1.2,-0.66)
{Baseline behavior;\\natural failure\\patterns};

\node[anchor=north west,font=\sffamily\bfseries\large,pal1!40!black,align=center,text width=5.4cm] at (5.3,-0.25) {111 findings};
\node[anchor=north west,font=\sffamily\large,align=center,text width=5.4cm,black] at (5.3,-0.85){High unsupported / hallucinated rate};

\node[toolbox, anchor=north west] at (14.7,-0.48) {};
\node[anchor=north west,font=\sffamily\bfseries\large,black] at (12.4,-0.21){Basic Detection Instruction};
\node[anchor=north west,font=\sffamily\large,text width=11.8cm,align=left,black] at (12.4,-0.6)
{Vulnerability detection request with CWE identification};
\node[toolbox, anchor=north west] at (14.7,-1.34) {};
\node[anchor=north west,font=\sffamily\bfseries\large,black] at (12.4,-1.37){Structured JSON Output};
\node[anchor=north west,font=\sffamily\large,text width=11.8cm,align=left,black] at (12.4,-1.76)
{Fixed output schema with no reasoning constraints};

\draw[connector,pal1!55!black](5.3,-0.72) -- (10.7,-0.72);
\draw[connector,pal1!55!black](-2,-2.6) -- (25.2,-2.6);

\begin{scope}[yshift=-2.6cm]
\filldraw[fill=pal4, draw=none](-2,-0.2) -- (3.5,-0.2) -- (4.5,-1.4) -- (3.5,-2.6) -- (-2,-2.6) -- cycle;
\filldraw[fill=pal4!45!black] (-1.6,-0.5) circle (0.36);
\node[text=white,font=\sffamily\bfseries\large] at (-1.6,-0.5) {v1};
\node[anchor=north west,font=\sffamily\bfseries\large,pal4!40!black] at (-1.2,-0.18){FEW-SHOT PROMPT};
\node[anchor=north west,align=left,font=\sffamily\large,text width=5.0cm,text=black] at (-1.2,-0.66)
{Hallucinated findings;\\unstable output\\structure};

\node[anchor=north west,font=\sffamily\bfseries\large,pal4!40!black,align=center,text width=5.4cm] at (5.3,-0.25){49 findings};
\node[anchor=north west,font=\sffamily\large,align=center,text width=5.4cm,black] at (5.3,-0.85){Noise reduced; pattern-based reasoning remained};

\node[toolbox, anchor=north west] at (14.7,-0.48) {};
\node[anchor=north west,font=\sffamily\bfseries\large,black] at (12.4,-0.21){Vulnerable / Non-vulnerable Examples};
\node[anchor=north west,font=\sffamily\large,text width=11.8cm,align=left] at (12.4,-0.6)
{Few-shot exemplars anchoring expected model behavior};
\node[toolbox, anchor=north west] at (14.7,-1.34) {};
\node[anchor=north west,font=\sffamily\bfseries\large,black] at (12.4,-1.37){Output \& Constraint Guidance};
\node[anchor=north west,font=\sffamily\large,text width=11.8cm,align=left] at (12.4,-1.76)
{Explicit reporting constraints reduce malformed responses};

\draw[connector,pal4!55!black](5.3,-0.72) -- (10.7,-0.72);
\draw[connector,pal4!55!black](-2,-2.6) -- (25.2,-2.6);
\end{scope}

\begin{scope}[yshift=-5.2cm]
\filldraw[fill=pal7, draw=none](-2,-0.2) -- (3.5,-0.2) -- (4.5,-1.4) -- (3.5,-2.6) -- (-2,-2.6) -- cycle;
\filldraw[fill=pal7!45!black] (-1.6,-0.5) circle (0.36);
\node[text=white,font=\sffamily\bfseries\large] at (-1.6,-0.5) {v2};
\node[anchor=north west,font=\sffamily\bfseries\large,white] at (-1.2,-0.18){EVIDENCE-BASED PROMPT};
\node[anchor=north west,align=left,font=\sffamily\large,text width=5.0cm,text=white] at (-1.2,-0.66)
{Unsupported claims;\\keyword-driven\\reasoning};

\node[anchor=north west,font=\sffamily\bfseries\large,pal7!40!black,align=center,text width=5.4cm] at (5.3,-0.25){37 findings};
\node[anchor=north west,font=\sffamily\large,align=center,text width=5.4cm,black] at (5.3,-0.85){Fewer false positives; more conservative};

\node[toolbox, anchor=north west] at (14.7,-0.48) {};
\node[anchor=north west,font=\sffamily\bfseries\large,black] at (12.4,-0.21){Source-to-Sink Reasoning};
\node[anchor=north west,font=\sffamily\large,text width=11.8cm,align=left] at (12.4,-0.6)
{Requires input, propagation, and sink evidence chain};
\node[toolbox, anchor=north west] at (14.7,-1.34) {};
\node[anchor=north west,font=\sffamily\bfseries\large,black] at (12.4,-1.37){Uncertainty Handling};
\node[anchor=north west,font=\sffamily\large,text width=11.8cm,align=left] at (12.4,-1.76)
{Permits rejecting findings lacking sufficient evidence};

\draw[connector,pal7!55!black](5.3,-0.72) -- (10.7,-0.72);
\draw[connector,pal7!55!black](-2.0,-2.6) -- (25.2,-2.6);
\end{scope}

\begin{scope}[yshift=-7.8cm]
\filldraw[fill=pal10, draw=none](-2,-0.2) -- (3.5,-0.2) -- (4.5,-1.4) -- (3.5,-2.6) -- (-2,-2.6) -- cycle;
\filldraw[fill=pal10!45!black] (-1.6,-0.5) circle (0.36);
\node[text=white,font=\sffamily\bfseries\large] at (-1.6,-0.5) {v3};
\node[anchor=north west,font=\sffamily\bfseries\large,white] at (-1.2,-0.18){FINAL REFINED PROMPT};
\node[anchor=north west,align=left,font=\sffamily\large,text width=5.0cm,text=white] at (-1.2,-0.66)
{CWE errors from\\artificial taxonomy\\constraints};

\node[anchor=north west,font=\sffamily\bfseries\large,pal10!35!black,align=center,text width=5.4cm] at (5.3,-0.32){26 findings};
\node[anchor=north west,font=\sffamily\large,align=center,text width=5.4cm,black] at (5.3,-0.92){Substantially more reliable findings};

\node[anchor=north west,font=\sffamily\bfseries\large,black] at (12.4,-0.06){Unrestricted CWE Reasoning};
\node[anchor=north west,font=\sffamily\large,text width=11.8cm] at (12.4,-0.46){Removed predefined Top-25 CWE scope restriction};

\node[anchor=north west,font=\sffamily\bfseries\large,black] at (12.4,-1.37){Corrected Inference Pipeline};
\node[anchor=north west,font=\sffamily\large,text width=11.8cm] at (12.4,-1.76){System prompt no longer concatenated with file content};

\draw[connector,pal10!45!black](5.3,-0.72) -- (10.7,-0.72);
\end{scope}

\filldraw[fill=white,draw=gray!40,rounded corners=0.25cm](-2,-10.7)rectangle(25.2,-11.55);
\node[anchor=west,font=\sffamily\bfseries\large,black] at (-2,-11.12){Derived via FDPR};
\node[anchor=west,font=\sffamily\large] at (12.4,-11.12){$\|$ Observation $\|$ Root Cause $\|$ Taxonomy $\|$ Prioritized Refinement};

\end{tikzpicture}
}\vspace{-2mm}
\caption{Prompt versions produced by FDPR. Each version's design decisions
are directly traceable to the dominant failure category identified in the
preceding refinement iteration; the outcome column reports the number of
manually validated findings on DVJA after each revision.}
\label{fig:prompt-versions}\vspace{-5mm}
\end{figure*}

\BfPara{Evaluation} 
We combine quantitative evaluation with qualitative failure analysis. Quantitative metrics measure vulnerability detection performance against ground-truth labels, while qualitative analysis assesses the reliability of model-generated reports.

\ItPara{Quantitative Evaluation} 
We report precision, recall, F1-score, and CWE classification accuracy using three matching criteria:\vspace{-2mm}

\begin{itemize}
    \item \textbf{Level 1 (Strict):} File, line number, and CWE category match.
    \item \textbf{Level 2 (Moderate):} File and CWE category match.
    \item \textbf{Level 3 (Relaxed):} CWE category match only.\vspace{-2mm}
\end{itemize}

Level~2 serves as the primary evaluation metric because source-to-sink vulnerabilities frequently span multiple lines while representing the same underlying weakness. During prompt refinement on DVJA, qualitative analysis guided prompt revisions, with quantitative results used to prioritize failure categories. External validation on the Juliet Test Suite uses Level~2 metrics, reported as Detection Rate and CWE Accuracy.

\ItPara{Qualitative Failure Analysis} 
Every reported vulnerability was manually reviewed to assess (i) correctness of the vulnerability claim, (ii) correctness of the assigned CWE, (iii) availability of supporting source-code evidence, (iv) consistency of the reasoning with the program logic, and (v) hallucinated or fabricated evidence. This analysis complements quantitative metrics by evaluating the reliability of model-generated security reports. Following each refinement iteration, incorrect findings were classified according to their underlying failure mechanism rather than their observable outcome. For example, reporting a vulnerability solely because a variable is named \texttt{password} is categorized as \emph{security-pattern over-generalization}, whereas inventing a nonexistent authentication or data-flow path is categorized as \emph{hallucinated evidence}. The resulting failure labels directly guided subsequent prompt refinements.

\section{Results and Failure Analysis}
\label{sec:results}

This section presents the results of the failure-driven prompt refinement process on DVJA and the external evaluation of the final prompt on the Juliet Test Suite.

\subsection{Prompt Refinement Outcomes}

Prompt refinement improved the reliability of model-generated reports by reducing unsupported findings, hallucinated evidence, and incorrect vulnerability reasoning. Table~\ref{tab:output-evolution} summarizes the evolution of model outputs across the four refinement iterations.

\begin{table}[t]
\centering
\caption{Evolution of generated findings during prompt refinement.}
\label{tab:output-evolution}\vspace{-2mm}
\small
\begin{tabularx}{\linewidth}{@{}l l c X@{}}
\toprule
\textbf{Version} &
\textbf{Refinement} &
\textbf{Findings} &
\textbf{Primary Observation} \\
\midrule
v0 & Baseline & 111 & Frequent unsupported and hallucinated findings \\
v1 & Improve reporting & 49 & Reduced noise; pattern-based reasoning remained \\
v2 & Evidence-based reasoning & 37 & Fewer false positives; more conservative analysis \\
v3 & Remove CWE restrictions & 26 & Most reliable and consistent results \\
\bottomrule
\end{tabularx}\vspace{-5mm}
\end{table}

Reported findings decreased from 111 in the baseline prompt (v0) to 26 in the final prompt (v3). This reduction reflects improved precision rather than lower detection capability. Manual analysis showed that successive refinements eliminated unsupported vulnerability claims, hallucinated evidence, and keyword-driven false positives, producing fewer but substantially more reliable security reports. 
Each refinement targeted the highest-priority failure category identified during the previous iteration. Early revisions focused on reducing unsupported findings, subsequent revisions strengthened evidence-based reasoning, and the final revision removed artificial constraints on CWE prediction. As the refinement progressed, obvious reasoning failures became less frequent, allowing the remaining errors to be analyzed at a finer level of granularity.

To complement the qualitative analysis, we measured recall across the ten OWASP Top~10 (2013) categories represented in DVJA. At baseline (v0), no individual model recovered all categories: qwen3-coder:30B identified 6/10 (60\%), qwen3.5:35B identified 8/10 (80\%), deepseek-coder:6.7B identified 0/10 (0\%), and llama3.2:3B identified 1/10 (10\%). The combined findings of all four models nevertheless covered all ten categories, albeit with the high false-positive rate reflected in the 111 reported findings.

After refinement (v3), the combined results covered all ten OWASP categories, but with zero false positives verified for each individual model. The 26 reported findings all corresponded to valid OWASP categories and frequently contained multiple correct CWE-level findings within a single category rather than spurious additional categories.

Manual inspection further showed that several apparent misses reflected evolution of the OWASP taxonomy rather than incorrect vulnerability reasoning. For example, Missing Function Level Access Control, Sensitive Data Exposure, and Cross-Site Request Forgery were reorganized under broader categories in later OWASP revisions~\cite{owasp2017top10,owasp2021top10}. When these taxonomy changes are taken into account, qwen3-coder:30B and qwen3.5:35B achieve taxonomy-adjusted recall of 10/10 (100\%). Applying the same analysis to deepseek-coder:6.7B and llama3.2:3B yields taxonomy-adjusted recall values of 4/10 (40\%) and 6/10 (60\%), respectively. We therefore report both strict recall, based on the original OWASP 2013 labels, and taxonomy-adjusted recall, which credits semantically equivalent classifications under later OWASP revisions. Because the analyses for deepseek-coder:6.7B and llama3.2:3B were performed at a lower level of granularity than those for qwen3-coder:30B and qwen3.5:35B, their taxonomy-adjusted recall values should be interpreted as preliminary.

We report only the v0 baseline and the v3 endpoint. Per-model recall was not tracked for the intermediate prompt versions, and only the zero-false-positive result was verified individually at v3, which we acknowledge as a limitation.

\subsection{Baseline Failure Analysis (v0)}

Manual analysis of the baseline outputs (v0) showed that most incorrect predictions are from a small number of recurring reasoning failures rather than isolated model mistakes. Across the evaluated models, three dominant failure categories consistently emerged: evidence hallucination, security pattern over-generalization, and incorrect CWE classification. These recurring failures guided the subsequent prompt refinement process.

\subsubsection{Evidence Hallucination}

The most severe baseline failure was the generation of reports unsupported by the analyzed source code. Models frequently inferred vulnerabilities from security-related identifiers or fabricated program behavior that did not exist. For example, variables named \texttt{password} were classified as hardcoded credentials despite the absence of assigned values, authentication logic, or credential exposure.

These errors indicate that the models attempted to satisfy the vulnerability detection objective without establishing sufficient evidence. This observation motivated the introduction of explicit evidence requirements, allowing the model to abstain when the available evidence was insufficient rather than forcing a vulnerability prediction.

\subsubsection{Security Pattern Over-Generalization}

A second recurring failure was excessive reliance on lexical security patterns. Models associated database operations with SQL injection, HTML output with cross-site scripting, and security-related variable names with credential exposure without verifying that the necessary exploitation conditions were present. 
These observations demonstrate that recognizing security-related concepts is not equivalent to identifying exploitable vulnerabilities. Consequently, later prompt revisions required explicit reasoning over the source, propagation path, sink, and sanitization before reporting a vulnerability. This refinement substantially reduced keyword-driven false positives in subsequent iterations.

\subsubsection{CWE Classification Errors}

The baseline analysis also showed that vulnerability identification and CWE classification are distinct reasoning tasks. Models often recognized the underlying security weakness while assigning an incorrect or overly general CWE identifier. Representative examples are summarized in Table~\ref{tab:cwe-baseline-errors}. These observations indicate that identifying a vulnerability and assigning the correct CWE require different reasoning capabilities. Consequently, prompt refinement treated vulnerability identification and CWE classification as separate stages rather than a single prediction task.

\begin{table}[t]
\centering
\caption{Representative reasoning errors during the baseline evaluation (v0).}
\label{tab:cwe-baseline-errors}\vspace{-2mm}
\small
\begin{tabular}{@{}N{3.0cm}N{6.1cm}@{}}
\toprule
\textbf{Observed Behavior} & \textbf{Incorrect Reasoning} \\
\midrule
Password variable &
Assumed hardcoded credentials (CWE-798) \\

Database operation &
Assumed SQL injection (CWE-89) \\

HTML output &
Assumed XSS without supporting evidence \\
\bottomrule
\end{tabular}\vspace{-3mm}
\end{table}

\subsection{Failure Taxonomy and Prompt Refinement}

The recurring failure mechanisms identified during the baseline analysis were consolidated into the failure taxonomy shown in Table~\ref{tab:taxonomy}. Rather than prioritizing failures solely by frequency, the taxonomy ranks them according to their impact on report reliability, analyst confidence, verification effort, and their suitability for mitigation through prompt refinement. 
The taxonomy served as the roadmap for prompt refinement. Each refinement iteration targeted the highest-priority failure category, translated its underlying cause into a corresponding prompt modification, and re-evaluated the updated prompt before addressing subsequent categories. This iterative process progressively eliminated dominant sources of unreliable behavior, enabling later iterations to focus on increasingly subtle reasoning and CWE classification errors.

\begin{table*}[t]
\centering
\caption{Failure taxonomy derived during prompt refinement.}
\label{tab:taxonomy}\vspace{-2mm}
\small
\scalebox{0.8}{\begin{tabular}{llll}
\toprule
\textbf{Failure Category} &
\textbf{Observed Behavior} &
\textbf{Underlying Cause} &
\textbf{Refinement Objective} \\
\midrule

Evidence hallucination &
Unsupported claims &
Missing evidence &
Require code evidence \\

Pattern over-gen. &
Keyword-based reports &
Lexical reasoning &
Require source-to-sink reasoning \\

CWE taxonomy errors &
Incorrect CWE assignment &
Weak hierarchy navigation &
Separate reasoning from classification \\

Over-reporting &
Report every suspicion &
No uncertainty handling &
Allow abstention \\

Taxonomy constraints &
Forced CWE mapping &
Restricted CWE scope &
Remove CWE restrictions \\

\bottomrule
\end{tabular}}\vspace{-3mm}
\end{table*}

\subsection{External Validation on the Juliet Test Suite}

To evaluate whether the refinement principles derived from DVJA generalize beyond the development dataset, the final prompt (v3) was evaluated on the Juliet Test Suite without further modification. Evaluation used preprocessed source code in which vulnerability comments, CWE identifiers, and identifying directory names were removed, ensuring that predictions were based solely on program behavior rather than benchmark metadata. Table~\ref{tab:juliet-config} summarizes the evaluation configuration.

\begin{table}[t]
\centering
\caption{Juliet evaluation configuration.}
\label{tab:juliet-config}\vspace{-2mm}
\small
\begin{tabular*}{\linewidth}{@{\extracolsep{\fill}}ll|ll@{}}
\toprule
\textbf{Component} & \textbf{Configuration} &
\textbf{Component} & \textbf{Configuration} \\
\midrule
Dataset & Juliet Test Suite (Java) &
Prompt & Final FDPR prompt (v3) \\

Models & Qwen3-Coder-30B; Qwen3.5-35B &
CWE scope & None \\

Preprocessing & Comments removed; file paths blinded &
Evaluation & File-level CWE classification \\
\bottomrule
\end{tabular*}\vspace{-3mm}
\end{table}

The evaluation processed 54,440 records, comprising 38,482 vulnerability predictions and 15,958 sentinel records. Sentinel records indicate files for which the model reported no vulnerability and are included only to verify complete dataset coverage; they carry no CWE label and are excluded from all reported performance metrics.

\subsubsection{Overall Performance}

Table~\ref{tab:overall-performance} summarizes the overall results. Both models maintained high CWE classification accuracy after applying FDPR. Although their detection rates differed, the models were evaluated on different subsets of Juliet due to computational constraints and should therefore not be compared directly. Instead, the results demonstrate that the refinement methodology transfers successfully to both a code-specialized model and a general-purpose LLM.

\begin{table}[t]
\centering
\caption{Overall model performance on the Juliet Test Suite with Predictions (Pr), Files Flagged (FF), Files Processed (FP), and Detection Rate (DR).}
\label{tab:overall-performance}\vspace{-2mm}
\small
\begin{tabular}{@{}lrrrrrr@{}}
\toprule
\textbf{Model} &
\textbf{Pr} &
\textbf{FF} &
\textbf{FP} &
\textbf{DR} &
\textbf{CWE Accuracy} &
\textbf{Unique CWEs} \\
\midrule
qwen3-coder:30B &
34,084 &
31,782 &
40,863 &
77.8\% &
97.0\% &
86 \\

qwen3.5:35B &
4,398 &
4,201 &
11,078 &
37.9\% &
93.4\% &
100 \\
\bottomrule
\end{tabular}\vspace{-3mm}
\end{table}

Across both models, 37,154 of 38,482 reported vulnerabilities received the correct CWE label, corresponding to an overall classification accuracy of 96.5\%. 



Classification accuracy was consistently high for vulnerability classes with well-defined semantic boundaries, including SQL injection, OS command injection, LDAP injection, HTTP header injection, divide-by-zero, and integer overflow. These results indicate that once unsupported reasoning and artificial taxonomy constraints were removed through FDPR, the models were able to assign CWE identifiers with high consistency across a broad range of vulnerability categories.

\subsection{Analysis of Remaining Errors}

Although the final prompt substantially reduced unsupported reasoning and unreliable vulnerability reports, manual analysis showed that the remaining errors were dominated by taxonomy navigation rather than vulnerability recognition. Of the 1,328 incorrect CWE predictions, most reflected selection of an incorrect level within the CWE hierarchy rather than failure to identify the underlying security weakness.

\subsubsection{Hierarchy Collapse}

The most common remaining error was \emph{CWE hierarchy collapse}, accounting for 978 misclassifications (73.6\%). In these cases, the models correctly identified the vulnerability family but predicted a broader parent CWE instead of the more specific child CWE expected by Juliet. Representative examples are shown in Table~\ref{tab:hierarchy-collapse}.

\begin{table}[t]
\centering
\caption{Representative examples of CWE hierarchy collapse.}
\label{tab:hierarchy-collapse}\vspace{-3mm}
\small
\begin{tabular}{@{}N{1.7cm}N{1.7cm}N{7.2cm}@{}}
\toprule
\textbf{Expected} & \textbf{Predicted} & \textbf{Interpretation} \\
\midrule
CWE-80 & CWE-79 & Generic XSS selected instead of the specific subtype \\
CWE-81 & CWE-79 & Error-message XSS mapped to its parent category \\
CWE-83 & CWE-79 & Attribute-based XSS mapped to its parent category \\
\bottomrule
\end{tabular}\vspace{-5mm}
\end{table}

Although incorrect under exact matching, these predictions show models generally recognized the underlying vulnerability while failing to navigate CWE granularity.

\subsubsection{Other Taxonomy Errors}

Credential-related errors accounted for 206 incorrect predictions (15.5\%), primarily reflecting confusion between CWE-259 (Hardcoded Password) and its parent CWE-798 (Hardcoded Credentials). Path-related errors accounted for 71 incorrect predictions (5.3\%), where specific variants such as CWE-23 and CWE-36 were mapped to the broader CWE-22. Together, hierarchy collapse, credential ambiguity, and path-related ambiguity accounted for 1,255 of the 1,328 incorrect predictions (94.5\%). The remaining 73 errors (5.5\%) were infrequent and did not exhibit a dominant shared failure mechanism. These results indicate that, after prompt refinement, the primary limitation shifted from vulnerability recognition to selecting the appropriate level of abstraction within the CWE taxonomy.

\subsubsection{Impact of Removing CWE Restrictions}

One of the final prompt refinements removed the predefined CWE scope imposed during earlier development iterations. Failure analysis showed that these restrictions occasionally forced correctly identified vulnerabilities into semantically related but incorrect CWE categories.

The Juliet evaluation validates this refinement. Without predefined CWE restrictions, the evaluated models correctly identified 86 and 100 unique CWE categories, respectively (Table~\ref{tab:overall-performance}), demonstrating that unrestricted taxonomy reasoning improves semantic classification across diverse vulnerability types.

\BfPara{Key Findings} 
The results lead to four primary observations. First, iterative prompt refinement reduced unsupported vulnerability reports, hallucinated evidence, and keyword-driven reasoning, producing fewer but more reliable security findings. Second, model failures clustered into a small number of recurring reasoning patterns that could be systematically addressed through targeted prompt refinements. Third, the resulting refinement principles generalized from the development dataset to the independent Juliet benchmark while maintaining high CWE classification accuracy. Finally, after refinement, the dominant remaining errors shifted from vulnerability recognition to CWE taxonomy navigation, where models more often selected an incorrect level of abstraction within the CWE hierarchy than failed to identify the vulnerability itself.
\section{Discussion}
\label{sec:discussion}

The objective of FDPR is not to produce a single optimized prompt, but to establish a systematic methodology for improving prompt reliability through failure analysis. Rather than treating model failures as isolated prediction errors, FDPR views them as engineering evidence that can guide prompt refinement. The experimental results show that recurring failure patterns can be identified, categorized, prioritized, and systematically translated into targeted prompt modifications, moving prompt engineering beyond empirical trial-and-error toward an evidence-driven engineering process.

\BfPara{Beyond Trial-and-Error} 
Current prompting typically relies on iterative experimentation: prompts are modified, evaluated, and compared using aggregate metrics such as accuracy, precision, recall, or F1-score. Although effective for selecting among alternatives, these metrics provide limited insight into why a prompt succeeds or fails. Two prompts may achieve similar performance while exhibiting fundamentally different failure behaviors, such as hallucinating unsupported vulnerabilities or conservatively missing difficult cases. 
FDPR instead treats failures as the primary refinement signal. Rather than asking which prompt performs better, it asks why the model failed and what prompt modification addresses that failure. Each refinement follows the same systematic process: identifying recurring failure patterns, analyzing their underlying causes, prioritizing them according to practical impact, and translating them into targeted prompt modifications. This shifts prompt engineering from empirical optimization toward a reproducible engineering methodology grounded in observable model behavior.

\BfPara{Failure Prioritization} 
A distinguishing characteristic of FDPR is explicit failure prioritization. Rather than addressing the most frequent failures first, FDPR ranks failure categories according to their practical impact on analyst trust, verification effort, reliability, security consequences, recurrence, and feasibility of prompt-based mitigation.

This prioritization proved particularly important during prompt refinement. Early prompt versions produced many unsupported vulnerability reports. Although these increased apparent vulnerability coverage, they substantially reduced practical usability because every unsupported finding required manual verification. Consequently, reducing hallucinated evidence and unsupported reasoning was prioritized before improving CWE classification accuracy. This reflects an important distinction between benchmark optimization and practical deployment: a system that reports more vulnerabilities is not necessarily more useful if its findings cannot be trusted.

\BfPara{Evolution of Failure Modes} 
The refinement process also revealed that model failures evolved as prompt quality improved. Early iterations were dominated by evidence hallucination, unsupported reasoning, and pattern-based false positives, indicating that the models frequently relied on superficial lexical associations rather than semantic program analysis. As these behaviors were progressively eliminated, the remaining errors shifted toward navigation of the hierarchical CWE taxonomy.

This transition is significant because it demonstrates that prompt refinement improved more than benchmark performance. The dominant remaining errors were no longer failures to recognize vulnerabilities, but failures to select the appropriate level of abstraction within the CWE hierarchy. Such errors are substantially more interpretable and bounded than hallucinated vulnerability reports, suggesting that FDPR transforms unreliable reasoning into more understandable and explainable failure modes.

\BfPara{Broader Implications} 
Although evaluated in the context of software vulnerability analysis, FDPR is not inherently specific to cybersecurity. Many LLM-based software engineering tasks, including code generation, automated repair, code review, and test generation, face similar challenges involving hallucinated outputs, unreliable reasoning, and limited explainability. Treating recurring failures as engineering evidence rather than isolated prediction errors provides a systematic methodology for improving prompt reliability across such applications. More broadly, this work suggests that prompt engineering should evolve from empirical prompt optimization into a reproducible engineering discipline. Rather than selecting among predefined prompting paradigms, reliable prompts may emerge by combining complementary prompting mechanisms guided by observed failure modes. In this view, evaluation metrics validate design decisions, while systematic failure analysis drives prompt evolution.

\section{Conclusion}
\label{sec:Conclusion}

We present FDPR, a failure-driven methodology for refining prompts for LLM-based vulnerability detection. By analyzing recurring failures and translating them into targeted prompt refinements, FDPR improved detection reliability while making remaining errors more interpretable. More broadly, the work demonstrates that failure analysis can serve as a principled foundation for prompt engineering beyond trial-and-error.

\end{document}